\documentclass[twoside,12pt]{article}
\newcommand{\be}{\begin{equation}}
\newcommand{\ee}{\end{equation}}
\date{ }
\begin{document}
\setcounter{equation}{0}
\setcounter{section}{0}
\title{ \bf Canonical Transformations and Squeezing in Quantum Mechanics}
\author {{\bf Jose M. Cerver\'o and Alberto Rodr\'{\i}guez} \\ {\small \bf
F\'{\i}sica Te\'orica}.
{\small \bf Facultad de Ciencias}.
{\small \bf Universidad de Salamanca}\\ {\small \bf 37008. Salamanca. Spain}}
\maketitle
\begin{abstract} In this Paper we present an approach to Quantum Mechanical
Canonical Transformations. Our main
result is that Time Dependent Quantum Canonical Transformations can always
be cast in the form of Squeezing
Operators. We revise the main properties of these operators in regard to
its Lie group properties, how two of
them can be combined to yield another operator of the same class and how
can also be decomposed and fragmented.
In the second part of the paper we show how this procedure works extremely
well for the Time Dependent Quantum
Harmonic Oscillator. The issue of the systematic construction of Quantum
Canonical Transformations is also
discussed along the lines of Dirac, Wigner and Schwinger ideas and to the
more recent work by Lee. The main
conclusion is that the Classical Phase Space Transformation can be
maintained in the operator formalism but
the construction of the Quantum Canonical Transformation is not clearly
related to the Classical Generating
Function of a Classical Canonical Transformation. We propose the road of
Squeezing Operators rather than
the old one attached to Quantum Operators constructed under the guideline
of the exponential of the
Classical Generating Function.

\end{abstract}
\vskip 0.4cm
$\qquad\qquad\qquad${\bf PACS Numbers: 03.65.Ca,  03.65.Fd  and  42.50.Dv}
\vskip 0.3 true in
\newpage
\quad{\bf \Large Introduction.}
\setcounter{equation}{0}
The purpose of this Paper can be simply stated: A Time Dependent Unitary
Operator $W(t)$ transforming a given
Hamiltonian $H(t)$ in another Hamiltonian $\tilde H(t)$ in the form:
\begin{eqnarray}
W(t) H(t) W^\dagger(t) - i\hbar W(t) \dot W^\dagger(t) &=& \tilde
H(t)\nonumber
\end{eqnarray}
can always be written as a combination of Squeezing Operators. The
operation above described is usually associated to a Time
Dependent Canonical Transformation (TDCT). If we succeed in convincing the
skeptical reader that a TDCT can always be realized
in the quantum formalism by means of a Squeezing Transformation (ST) and
reciprocally, so that:
\begin{eqnarray}
TDCT \Longleftrightarrow ST \nonumber
\end{eqnarray}
then the rest of the Paper will merely be algebra and rethoric.

The subject of Quantum Canonical Transformations was initiated by Dirac
(Refs. \cite{DIRAC1} and \cite{DIRAC2}) and
subsequently developed by Wigner \cite{WIGNER1} and Schwinger \cite{SCHWI}.
In all these cases however the
formalism was mainly constrained to the Time Independent case. Much more
recently, the work of Lee and
collaborators (Refs. \cite{LEE1},
\cite{LEE2} and \cite{LEE3}) has added quite a lot of steam to the subject
but mainly with an eye on laying the foundations of
an unambiguous Quantum Hamilton-Jacobi Formalism. In the same spirit one
should also consider the work of Lewis and
collaborators \cite{LEWIS} whose goal seems to be addressed to the
construction of the Quantum analog to the Classical
Action-Angle Formalism. Although our aim has a different motivation it
turns out to share much more than we previously thought
with the above mentioned approaches.

This Paper is divided into two parts. Firstly we shall review the
properties of the Squeezing Operators keeping in mind that
excellent reviews are already available in the literature. This is why we
shall emphasize mainly only two groups of properties
of these Operators: Fragmentation and Multiplication or in a more
colloquial way to describe it: breaking them and gluing them
back together.

In the Second part we shall be discussing the nature of the
$W(t)$-operators. The aim is to show that every $W(t)$-operator
can be constructed by a simple rule of multiplication of Squeezing
Operators. At this point we shall need a great deal of the
properties obtained in the previous Section. The main example for
illustrating the procedure will be the Time Dependent
Harmonic Oscillator and many of the manipulations will be carried out for
the benefit of the reader using this physical system.
As it has been mentioned previously one may need to use some already
existing reviews and previously known results. For the
Squeezing Operators \cite{DAN} we recommend the review by Teich and Saleh
\cite{BAHA} among many other existing excellent
papers with similar tutorial approach and contents. The Time Dependent
Harmonic Oscillator has been treated throughout in
Reference
\cite{CERLEJA1} and many of the properties we shall be using can be found
either in this last Reference or in the more recent
account of Reference \cite{CERVERO}.
\section{Squeezing Operators}

Let the $SU(1,1)$ Lie Algebra be defined through its commutation relations:
\begin{eqnarray}
[K_+, K_-] = -2  K_o  \qquad\qquad\qquad [K_o, K_{\pm}] =  \pm K_{\pm}
\end{eqnarray}
A Squeezing Operator shall be defined henceforth in the following general way:
\begin{eqnarray}
S(\theta, \rho) &=& \exp\{2i\theta(t) K_o\} \exp\{\rho(t) K_+ -\rho^*(t) K_-\}
\end{eqnarray}
One should emphasize at this point that this construction can also be
generalized to $SO(3)$ or any other Lie Algebra
\cite{CERLEJA2}. However we shall be considering just the $SU(1,1)$ case
for reasons which will become clear just below.
Actually the most popular and practical realization of the $SU(1,1)$ Lie
Algebra is the one given by the creation and
annihilation operators  of the Canonical Algebra by means of the obvious
identification:
\begin{eqnarray}
K_+ = {1\over2}{a^+}^2\qquad\qquad\qquad   K_- = {1\over2}{a}^2
\qquad\qquad\qquad    K_o =  {1\over2}(a^+ a + {1\over2})
\end{eqnarray}
Aside from this particular realization of the $SU(1,1)$ Lie Algebra, we
list some of the main properties of the Squeezing
Operators as defined in (1.2). Firstly we note that the following
transposition property holds:
\begin{eqnarray}
\exp\{2i\theta K_o\}\exp\{\rho K_+ -\rho^* K_-\} = \exp\{\rho e^{2i\theta}
K_+ -\rho^* e^{-2i\theta} K_-\} \exp\{2i\theta
K_o\}
\end{eqnarray}
A very important piece of information is the way in which these operators
can be multiplied giving rise to another operator of
the same kind. This is an obvious consequence of the group law. The first
crucial relationship can be written as:
\begin{eqnarray}
\exp\{\rho_2 K_+ -\rho_2^* K_-\}\exp\{\rho_1 K_+ -\rho_1^* K_-\}  &=&
\exp\{2i\theta_o K_o\}\exp\{\rho_o K_+
-\rho_o^* K_-\}
\end{eqnarray}
After stablishing the one to one correspondence:
\begin{eqnarray}
\rho_1  =  r_1 \exp\{i\varphi_1\}\qquad=>\qquad \eta_1  =  \tanh \{r_1\}
\exp\{i\varphi_1\}\\
\rho_2  =  r_2 \exp\{i\varphi_2\}\qquad=>\qquad \eta_2  =  \tanh \{r_2\}
\exp\{i\varphi_2\}
\end{eqnarray}
one can obtain $\theta_o$ and $\rho_o$ in the form:
\begin{eqnarray}
\eta_o &=& {{\eta_1 + \eta_2}\over {1 + \eta_1^*\eta_2}} =  \quad
\mid\eta_o\mid \exp\{ i\arg \eta_o\}
\end{eqnarray}
\begin{eqnarray}
\rho_o &=& \rm argtanh\{\mid\eta_0\mid\} \exp\{i \arg\eta_o\}
\end{eqnarray}
\begin{eqnarray}
2i\theta_o &=& \log\{{1 + \eta_1^*\eta_2\over 1 + \eta_1\eta_2^*}\}
\end{eqnarray}
Another useful group of properties are those having to do with
"fragmentation". Let us take the Squeezing Operator given by
(1.2). One can easily show that it can also be written as \cite{GERRY}:
\begin{eqnarray}
S(\theta(t), \rho(t)) = \exp\{2i\theta(t) K_o\} \exp\{\rho(t) K_+
-\rho^*(t) K_-\} =\nonumber \\
= \exp\{2i\theta(t) K_o\}\exp\{\eta(t)K_+\}\exp\{\gamma(t)
K_o\}\exp\{-\eta(t) K_-\}
\end{eqnarray}
where $\eta(t)$ and $\gamma(t)$ are given by correspondences of similar
sort of those listed in (1.6) and (1.7), namely:
\begin{eqnarray}
\rho(t)  =  r(t) \exp\{i\varphi(t)\}=> \eta(t)  =  \tanh \{r(t)\}
\exp\{i\varphi(t)\}
\end{eqnarray}
\begin{eqnarray}
\gamma(t)   = \ln(1-\mid \eta(t)\mid^2)
\end{eqnarray}
Finally the transformation of the Lie Algebra generators under the
Squeezing Operators are also extremely useful for some of
the calculations below presented:
\begin{eqnarray}
S(\theta, \rho) K_o S^{\dagger}(\theta, \rho) &=& \cosh\{2r(t)\}K_o -
{1\over2}\sinh\{2r(t)\}\{e^{i(\theta_+ +\theta_-)}K_+ + e^{-i(\theta_+
+\theta_-)}K_-\}\qquad\qquad\qquad
\end{eqnarray}
\begin{eqnarray}
S(\theta, \rho) K_+ S^{\dagger}(\theta, \rho) = \cosh^2\{r(t)\}
e^{2i\theta_+}K_+
+\sinh^2\{r(t)\} e^{-2i\theta_-}K_- - e^{i(\theta_+ -\theta_-)}
\sinh\{2r(t)\} K_o\qquad
\end{eqnarray}
\begin{eqnarray}
S(\theta, \rho) K_- S^{\dagger}(\theta, \rho) = \sinh^2\{r(t)\}
e^{2i\theta_-}K_+
+\cosh^2\{r(t)\} e^{-2i\theta_+}K_- - e^{-i(\theta_+ -\theta_-)}
\sinh\{2r(t)\} K_o\qquad
\end{eqnarray}
\vskip0.5cm
\noindent where
\begin{eqnarray}
\theta_+  =  \theta(t)  \quad\quad\qquad and \quad\quad\qquad \theta_-  =
\varphi(t) + \theta(t)
\end{eqnarray}
\section{Time Dependent Canonical Transformations.}
Suppose we start with a Time dependent Hamiltonian and we apply a sequence
of Time Dependent Canonical
Transformations in such a way that we go from the initial $H(t)$ to the
final $H_o(t)$ through the following set
of TDCT
\begin{eqnarray}
H(t)\qquad=>\qquad H_1(t)\qquad=>\qquad H_{12}(t) = H_o(t)\nonumber
\end{eqnarray}
This sequence really  means in terms of the actual application of the
$W(t)$-operators the following set of
step by step transformations:
\newpage
\begin{eqnarray}
W_1(t) H(t) W_1^\dagger(t) - i\hbar W_1(t) \dot W_1^\dagger(t) &=&  H_1(t)
\end{eqnarray}
\begin{eqnarray}
W_2(t) H_1(t) W_2^\dagger(t) - i\hbar W_2(t) \dot W_2^\dagger(t) &=&
H_{12}(t) = H_o(t)
\end{eqnarray}
There must be an operator which does the same job in just one step, namely:
\begin{eqnarray}
W(t) H(t) W^\dagger(t) - i\hbar W(t) \dot W^\dagger(t) &=&  H_o(t)
\end{eqnarray}
This operator must neccessarily be $W(t) = W_2(t)W_1(t)$ as one can easily
check by merely using the Time
Dependent Schr\"odinger Equation. This obvious fact suggests a group law.
Let this group be $SU(1,1)$. As a
general example for quadratic Time Dependent Hamiltonians we choose $H(t)$
to be:
\begin{eqnarray}
H(t) = {1\over 2m}\{ {\beta_3(t)}\hat p^2 +  \beta_2(t){m\omega_o}[\hat
x\hat p+\hat p\hat x]  + \beta_1(t) m^2 \omega_o^2 \hat x^2\}
\end{eqnarray}
And we choose as $W_1(t)$ and $W_2(t)$ the following unitary operators
\cite{CERLEJA1}:
\begin{eqnarray}
W_1(t) &=& \exp \{{i\over4\hbar}(\ln\beta_3(t))[\hat x\hat p+\hat p\hat
x]\}\\
W_2(t) &=& \exp \{{im\over2 \hbar}(\omega_o\beta_2(t) - {\dot
\beta_3(t)\over 2 \beta_3(t)}) \hat x^2\}
\end{eqnarray}
The resulting Hamiltonians in each of the steps (2.18) and (2.19) turn out
to be:
\begin{eqnarray}
H_1(t) = {\hat p^2\over 2m} + {1\over2}\{\omega_o\beta_2(t) -{\dot
\beta_3(t)\over 2 \beta_3(t)}\}[\hat x\hat
p+\hat p\hat x]  + {1\over2}m\omega_o^2\beta_1(t)\beta_3(t)\hat x^2
\end{eqnarray}
\begin{eqnarray}
H_{12}(t) = H_o(t) = {\hat p^2\over 2m}  + {1\over2}m\{\omega_o^2
(\beta_1\beta_3 -\beta_2^2) +
\omega_o({{\dot\beta_3\beta_2 - \dot\beta_2\beta_3}\over \beta_3}) +
{{\ddot\beta_3\over 2\beta_3}} - {{3\dot\beta_3^2\over 4\beta_3^2}}\}\hat
x^2\quad
\end{eqnarray}
The next step is to show that $W_1(t)$ and $W_2(t)$ can be written in a
Squeezing Operator form. It is not
hard to see that $W_1(t)$ can be written as:
\begin{eqnarray}
W_1(t) = \exp\{\rho_1(t) K_+ -\rho_1^*(t) K_-\}
\end{eqnarray}
where we have used the well known form of the creation and annihilation
operators as linear combinations of the
canonical operators $\hat x$ and $\hat p$. The function $\rho_1(t)$ has no
complex phase and turns out to be in
this particular case just a real function of the form:
\begin{eqnarray}
\rho_1(t)  &=& \rm argtanh\{{{1-\beta_3(t)}\over{1+\beta_3(t)}} \}
\end{eqnarray}
Furthermore the operator $W_2(t)$ can be written as:
\begin{eqnarray}
\nonumber W_2(t) &=& \exp \{i\gamma(t) [K_+  +  K_- + 2K_o]\} =
\end{eqnarray}
\begin{eqnarray}
= \exp\{2i\theta_2(t) K_o\}\exp\{\rho_2(t) K_+ - \rho_2^*(t) K_-\}
\end{eqnarray}
where
\begin{eqnarray}
\gamma(t)  &=& {1\over2\omega_o}\{\omega_o \beta_2(t) - {\dot\beta_3(t)
\over2\beta_3(t)}\}\\
\theta_2(t) &=& \arctan\gamma(t)\\
\rho_2(t)  &=& \rm argtanh \{{\gamma(t) \over (1 +
\gamma^2(t))^{1\over2}}\} \exp\{i ({\pi\over2} -
\theta_2(t))\}
\end{eqnarray}
Obviously (2.26) and (2.28) are Squeezing Operators as the one generaly
defined in (1.2) and can be written in
this notation as $S(0,\rho_1(t))$ and $S(\theta_2(t),\rho_2(t))$. It is now
just a matter of a tedious but
trivial calculation to combine them by making use of the expressions given
in Section 1. What we try to do is
basically to use the expressions (1.5)-(1.10) in order to obtain a single
Squeezing Operator in the form:
\begin{eqnarray}
\nonumber W(t) = W_2(t)W_1(t) &=&  \exp\{2i\theta_2(t) K_o\}
\exp\{\rho_2(t) K_+ -\rho_2^*(t) K_-\}
\exp\{\rho_1(t) K_+ -\rho_1^*(t) K_-\} \\   &=& \exp\{2i\theta_o  K_o \}
\exp\{\rho_o(t) K_+
-\rho_o^*(t) K_-\}
\end{eqnarray}
where in this case:
\begin{equation}
\rho_o(t)  ={\rm argtanh}\{\sqrt{{4\gamma^2(t) + (1-\beta_3(t))^2  \over
4\gamma^2(t) + (1+\beta_3(t))^2}}\}
\exp\{i({\rm arctan}\{{2\gamma(t)\over1-\beta_3(t)} \} -{\rm
arctan}\{{2\gamma(t) \over 1+\beta_3(t)}\})\}
\end{equation}
\begin{eqnarray}
\theta_o(t) &=& \arctan \{{2\gamma(t) \over 1 + \beta_3(t)} \}
\end{eqnarray}
In the final part of this Section we shall be discussing the relationship
of what so far has been done to
relate Squeezing Operators and Time Dependent Quantum Canonical
Transformations with what is believed to be
the standard lore on the connection between Quantum and Classical Canonical
Transformations. As it has been
previously mentioned the beginning of this sort of discussions is due to
Dirac, Wigner and Schwinger
(See References \cite{DIRAC1}, \cite{DIRAC2}, \cite{WIGNER1} and
\cite{SCHWI}). Recently a renewed interest
on the subject has been put forward by Lee and collaborators (\cite{LEE1},
\cite{LEE2} and \cite{LEE3}). Also
we should mention that quite recently Kim and Wigner \cite{WIGNER2}
mentioned the relationship between
squeezing and canonical transformations in the context of the time
independent quantum problem, cleverly relating
the squeezing properties to 2+1-Lorentz transformations due to the well
known fact that $SO(2,1)$ (The Lorentz
group in "flatland") is locally isomorphic to $SU(1,1)$. In the rest of
this paper we shall be dealing with
the Classical Time Dependent Canonical Transformations of the Hamiltonian:
\begin{eqnarray}
H(t) = {1\over 2m}\{ {\beta_3(t)}p^2 +  \beta_2(t){m\omega_o}[xp+ px]  +
\beta_1(t) m^2\omega_o^2 x^2\}
\end{eqnarray}
which is the Classical counterpart of (1.2). It is a trivial exercise in
Classical Mechanics
\cite{GOLDSTEIN} to find a Generating Function of a Canonical
Transformation in Phase Space which leads
(2.35) to
\begin{eqnarray}
H_o(t) = {1\over 2m} \{P^2 + m^2 \Omega^2(t) X^2\}
\end{eqnarray}
where $\Omega^2(t)$ is given by the expression:
\begin{eqnarray}
\Omega^2(t) &=& \omega_o^2\{\beta_1\beta_3 -\beta_2^2\} +
\omega_o\{{{\dot\beta_3\beta_2 - \dot\beta_2\beta_3}\over \beta_3}\} +
\{{1\over2}({{\ddot\beta_3\over \beta_3}} - {{\dot\beta_3^2\over
\beta_3^2}}) - {1\over4}({{\dot\beta_3^2\over \beta_3^2}})\}
\end{eqnarray}
which obviously coincides with the one appearing in (2.25).  In Classical
Mechanics a Generation Function
of the class two: $F_2^C(x,P,t)$ can be constructed such that:
\begin{eqnarray}
{\partial F_2^C(x,P,t)\over \partial x} = p   \qquad;\qquad {\partial
F_2^C(x,P,t)\over \partial P} = X
\qquad;\qquad H_o(t) =  H(t) + {\partial F_2^C(x,P,t) \over \partial t}\qquad
\end{eqnarray}
The explicit form of $F_2^C(x,P,t)$ in our case, takes the form:
\begin{eqnarray}
F_2^C(x,P,t) &=& {m\over{2 \beta_3(t)}}\{{\dot\beta_3(t)
\over2\beta_3(t)}-\omega_o \beta_2(t)\} x^2 +
\beta_3(t)^{-{1\over2}} x P
\end{eqnarray}
giving rise to the following canonical phase space transformation:
\begin{eqnarray}
X &=& \beta_3(t)^{-{1\over2}} x
\end{eqnarray}
\begin{eqnarray}
P &=& {\beta_3(t)}^{1\over2} \{p  + ({m \over
\beta_3(t)})(\omega_o\beta_2(t) - {\dot\beta_3(t)
\over 2\beta_3(t)}) x\}
\end{eqnarray}
The main question arises as to whether these Classical Phase Space
transformations which are under the basis
of a Canonical Transformations have anything to do with the Squeezing
Formalism. The idea of constructing the
Quantum Canonical Operator by merely using the Generating Function as Dirac
suggested (\cite{DIRAC1} and
\cite{DIRAC2}) is simply too naive and does not work: we cannot reproduce
$W(t)$ given by (2.32) (together
with (2.33) and (2.34)) just by constructing the dimensionless exponential
of the Classical Generating Function
(2.39):
\begin{eqnarray}
exp\{{i\over \hbar}F_2^C(x,P,t)\} \quad \neq\quad W(t)
\end{eqnarray}
With this result in mind one is tempted to conjecture that a Quantum
generating Function $F_2^Q(x,P,t)$ could be
defined in the same spirit as the Feynman Quantum Action used to construct
the Quantum Propagator. Such
$F_2^Q(x,P,t)$  should retain some features of the $F_2^C(x,P,t)$, in
particular the classical limit which
has to be properly defined. It is also encouraging that although these
Generating Functions obviously differ,
our Time Dependent Quantum Canonical Operator $W(t)$ yields the Phase Space
Transformation (2.40)-(2.41)
which is clearly Classical in origin but can be written in the language of
Canonical Operators. In fact one can
can actually check that the following relationships hold:
\begin{eqnarray}
W^\dagger(t)\hat x W(t) = \hat X =  \beta_3(t)^{-{1\over2}} \hat x
\end{eqnarray}
\begin{eqnarray}
W^\dagger(t)\hat p W(t) = \hat P = {\beta_3(t)}^{1\over2} \{\hat p  + ({m
\over \beta_3(t)})
(\omega_o\beta_2(t) - {\dot\beta_3(t) \over 2\beta_3(t)}) \hat x\}
\end{eqnarray}
The conditions expressed by the above equations have been already stressed
by Lee (\cite{LEE2} and \cite{LEE3})
as the actual main requirements for a Quantum Canonical Transformations
rather than the dubious association
(2.42) which is actually false. We see in this Time dependent analysis that
the Classical Phase Space
transformation still survives in the quantum Operator domain. To go further
we conjecture that the Squeezing
Operator Formalism will be much more fruitful in view of the results hereby
presented and an extension of these
will be presented in a future Report now in preparation.
\vskip 1cm
\noindent {\bf  Acknowledgment.} We acknowledge with thanks the support
provided
by the Research in Science and Technology Agency of the Spanish Government
({\bf DGICYT})
under contract {\bf PB98-0262}.

\end {document}